\documentclass[preprint]{jpsj2}
\usepackage{epsfig}

\def\runtitle{
Riddled-like Basin in Two-Dimensional Map
...
}
\def\runauthor{
Shohei {\sc Fukano},
Yumino {\sc Hayase}
and Hiizu {\sc Nakanishi}
}

\title
{
Riddled-like Basin in Two-Dimensional Map
for Bouncing Motion of an Inelastic Particle on a Vibrating Board
}

\author
{ 
Shohei {\sc Fukano}\footnote{E-mail: fukano@stat.phys.kyushu-u.ac.jp},
Yumino {\sc Hayase}\footnote{E-mail: yumino@stat.phys.kyushu-u.ac.jp},
and Hiizu {\sc Nakanishi}\footnote{E-mail: naka4scp@mbox.nc.kyushu-u.ac.jp} 
}

\inst
{
Department of Physics, Kyushu University 33, Fukuoka 812-8581
}

\recdate
{
March 22, 2002
}

\abst { Motivated by bouncing motion of an inelastic particle on a
vibrating board, a simple two-dimensional map is constructed and its
behavior is studied numerically.  In addition to the typical route to
chaos through a periodic doubling bifurcation, we found peculiar
behavior in the parameter region where two stable periodic attractors
coexist.  A typical orbit in the region goes through chaotic motion for
an extended transient period before it converges into one of the two
periodic attractors.  The basin structure in this parameter region is
almost riddling and the fractal dimension of the basin boundary is
close to two, {\it i.e.}, the dimension of the phase space.
}

\kword
{
transient chaos, boundary crisis, riddled basin,
bouncing inelastic particle, granular material
}

\begin{document}
\sloppy
\maketitle

\noindent

Particle motion accelerated by a vibrating boundary has been studied for
some time as an example of simple physical systems that show chaotic
behavior\cite{LL72,LLbook} since Fermi proposed a novel mechanism of
acceleration of a charged particle through an oscillating field (Fermi
acceleration)\cite{F49}.  A variant of the system is a bouncing
inelastic particle on a vibrating board, and this has also been studied,
particularly in connection with granular systems.  It
has been demonstrated experimentally\cite{P83,PKF85,KFP88} and
theoretically\cite{E86,L-A90,ML90,LM93} that even a single particle
bouncing on a board shows a variety of behaviors such as mode-locked
periodic motion, periodic doubling, chaotic behavior and quasi-periodic
motion.

In this letter, based on the observation of dynamics on a single
inelastic particle motion under gravity driven by a vibrating board,
we construct a simple two-dimensional map, and demonstrate numerically
that the map shows peculiar behavior, {\it i.e.}, a very long transient chaos
and a riddled-like basin structure in the parameter region where two
periodic attractors coexist.


The original system we look at in constructing our map is an inelastic
particle under the gravity; the particle bounces vertically on a
vibrating board with a restitution constant $r$.  Then, the particle
velocity just before the $n$'th bounce $u_{n}$ and the velocity just
after the bounce $v_{n}$ are related by
\begin{equation}
 -{v_{n}-V(t_n) \over u_{n}-V(t_n) } = r ,
\end{equation}
where $t_n$ is the time at the $n$'th bounce and $V(t)$ is the velocity
of the board at time $t$ (Fig.1).  If the motion of the board $V(t)$ is
given, we can determine, in principle, the time $t_{n+1}$ and the
particle velocity $u_{n+1}$ of the $n+1$'th bounce from $t_n$ and
$v_{n}$, and we should be able to obtain a map that describes the
particle motion by using this model. It is, however, impossible in general
to express the map in an explicit form\cite{LM93} because the board is
moving.  Here, we employ the following simplification that allows us to
construct the explicit map by which we can calculate $(v_{n+1},
t_{n+1})$ from $(v_{n}, t_{n})$.

First, we ignore the difference in the position of the board at each
bounce; then we have
\begin{equation}
 u_{n+1}  =  -v_{n},
\qquad
 t_{n+1} = t_n + {2\over g} v_{n}, 
\end{equation}
where $g$ denotes the gravitational acceleration.  This simplification
should be justified when the amplitude of the vibration is small
compared with the bouncing height (high bounce approximation).  Under
this simplification, the velocity just after the bounce must be
positive such that $t_{n+1}>t_n$.  To ensure this, we must assume that
the vibrating board always moves upward, {\it i.e.}, $V(t)\geq 0$.  This
appears to contradict the first simplification, but we may consider
that the board shifts downward discontinuously at each period.  In the
present work, we employ
\begin{equation}
 V(t) = A(\sin \omega t + 1), 
\end{equation}
then we have
\begin{equation}
\left\{
\begin{array}{rcl}
v_{n+1}  & = & rv_{n} + (r+1) A (\sin\omega t_{n+1} + 1)
\\
t_{n+1} & = & t_n + {\displaystyle 2\over\displaystyle g} v_{n}.
\end{array}\right.
\end{equation}

After rescaling the variables by the transformation,
\begin{equation}
v_n \to A v_n, \quad
t_n \to {A\over g}t_n,\quad
\omega \to {g\over A}\omega ,
\end{equation}
we have the two-dimensional map for $v_n$ and $t_n$
\begin{equation}
\left\{
\begin{array}{rcl}
v_{n+1} & = & rv_n + (1+r)(\sin\omega t_{n+1}+1)
\\
t_{n+1} & = & t_n + 2v_n
\end{array}\right.
\end{equation}
with the two parameters $r$ and $\omega$.
This was found to be equivalent to the one studied by
Everson\cite{E86}, but with different parameterization.


This map shows a typical route to chaos via periodic doubling
from the mode-locking bounce with the same period with the external
vibration as we change the parameter $\omega$ from 0.9 to 1.8 by
fixing $r=0.39$ (Fig.2a).  This periodic doubling and chaotic behavior
have previously been reported for the inelastic bounce system in
numerical\cite{E86,L-A90} and also experimental\cite{P83,PKF85,KFP88}
studies.  The strange attractor for the chaotic motion at $\omega=1.54$
and $r=0.39$ is shown in Fig.2b, which indicates that the attractor has
a fractal structure with a Housdorff dimension larger than 1 in the
two-dimensional phase space.

Figure 2(b) also shows a period one attractor and its basin;
this period one motion is locked to half the vibration frequency.
Upon changing the parameters further, 
this basin moves towards the chaos attractor and causes sudden
destruction of the chaos through a boundary crisis when the basin
touches the chaos attractor, then the chaos disappears as a true attractor,
but its vestige remains as a transient one.

After the destruction of the chaos, the period one attractor bifurcates
into period two, while a new period ten attractor is stabilized out of
the transient chaos and coexists with this period two attractor.  In
Fig.3, both attractors are shown in the phase space as well as the
``temporal attractor'' of the transient chaos which remains after the
crisis.  As is seen in this figure, the period ten attractor is embedded
in the transient attractor of the chaos.  In this parameter region,
where the period two and the period ten orbits coexist, the system
shows peculiar behavior, as we describe below.

First, the system goes through an extremely long transient period
before the trajectory converges
on one of the two periodic orbits, if the initial state is chosen
outside the direct basin of the period two attractor.
The length of the transient easily exceeds one thousand iterations.
The reason for this unusually long transient is that the 
region of direct attraction around the period ten attractor is extremely 
small.


Second, the basin structures of these attractor are scattered and mixed
each other almost completely (Fig.4).  To characterize the structure of
the basins, the uncertainty exponent, or final state sensitivity,
$\alpha$ is introduced as the exponent for the probability $f(\epsilon)$
that two points separated from each other by the distance $\epsilon$ in
the phase space are in different basins:
\begin{equation}
f(\epsilon) \sim \epsilon^\alpha .
\end{equation}
The small value of $\alpha$ implies that the two basins are mixed each other
rather wildly.  The probability $f(\epsilon)$ is plotted in Fig.5 for
$\omega=1.5852$ and $r=0.39$, and the estimated value of
$\alpha$ is $6.2\times 10^{-4}$.  This means the fractal dimension $d_f$
of the basin boundary is very close to 2 since the uncertainty exponent
$\alpha$ is related to $d_f$ by
\begin{equation}
 d_f = d - \alpha ,
\end{equation}
where $d$ is the spatial dimension of the phase space.\cite{LLbook}


The typical appearance of the basin structure of the period ten
attractor in Fig.4 strongly resembles the riddled basin, {\it i.e.}, a
basin that is riddled with holes.\cite{PG91,OAKSY94}
Mathematically, the riddled basin is characterized by the following two
features: (i) it has nonzero measure, (ii) any open neighbor of a point
in the basin contains a point that does not belong to the basin.  Such a
basin has been found for a chaos attractor that is confined within a
subspace of the whole phase space due to the symmetry of the system.
According to this definition, a riddled basin is not an open set,
therefore the present case cannot be a riddled basin; the basins for
both the period two and the period ten attractors are open because the
attractors have a finite neighbor around them that is contained within
the basin, therefore whole basins can be constructed by the union
of infinite pre-images of these neighbors.

On the other hand, the basin shown in Fig.4 cannot be
distinguished from a riddled basin by almost any direct measure that
can be calculated from the geometry of the basin.  Any numerical
evidence that the present basin is an open set is wiped away by the
fragmentation of the pre-images during the extremely long lifetime of
the transient chaos.  This situation, however, does not seem to be
unusual for the periodic window in the chaos; in such
cases, the unstable periodic orbit embedded in the chaos attractor is
stabilized extremely weakly, as in the present case.

In summary, we have examined a simple two-dimensional map that describes
an inelastic ball bouncing on a vibrated board, and found that the
system shows a riddled-like basin structure.  The basin should not be a
riddled basin based on the mathematical definition, but its uncertainty
exponent is very close to 2 and it almost cannot be distinguished from a
riddled basin.  This is due to the very long lifetime of the transient
chaos.  This situation, however, seems to be generic when
the unstable periodic orbit in the chaos attractor is stabilized.
Note that the mechanism does not require any symmetry in the system in
contrast with the case of the riddled basin.

The authors are grateful to Professor Kaneko for his useful comments.


\newpage

\centerline{Shohei {\sc Fukano},Yumino {\sc Hayase},and Hiizu {\sc Nakanishi}}
\vskip 3.5cm
\begin{figure}
\begin{center}
\epsfig{width=10cm,file=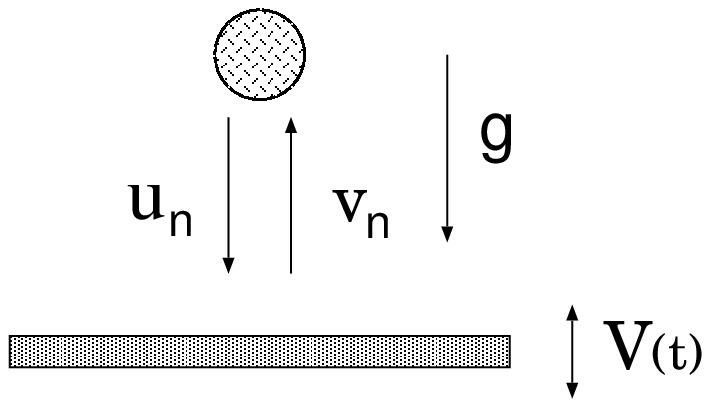}
\end{center}
\vskip 3.5cm
\caption{An inelastic particle bouncing on a vibrating board.}
\label{Fig-1}
\end{figure}
\newpage

\setcounter{figure}{1}
\centerline{Shohei {\sc Fukano},Yumino {\sc Hayase},and Hiizu {\sc Nakanishi}}
\vskip 2.cm
\begin{figure}
\begin{center}
\epsfig{width=10cm,file=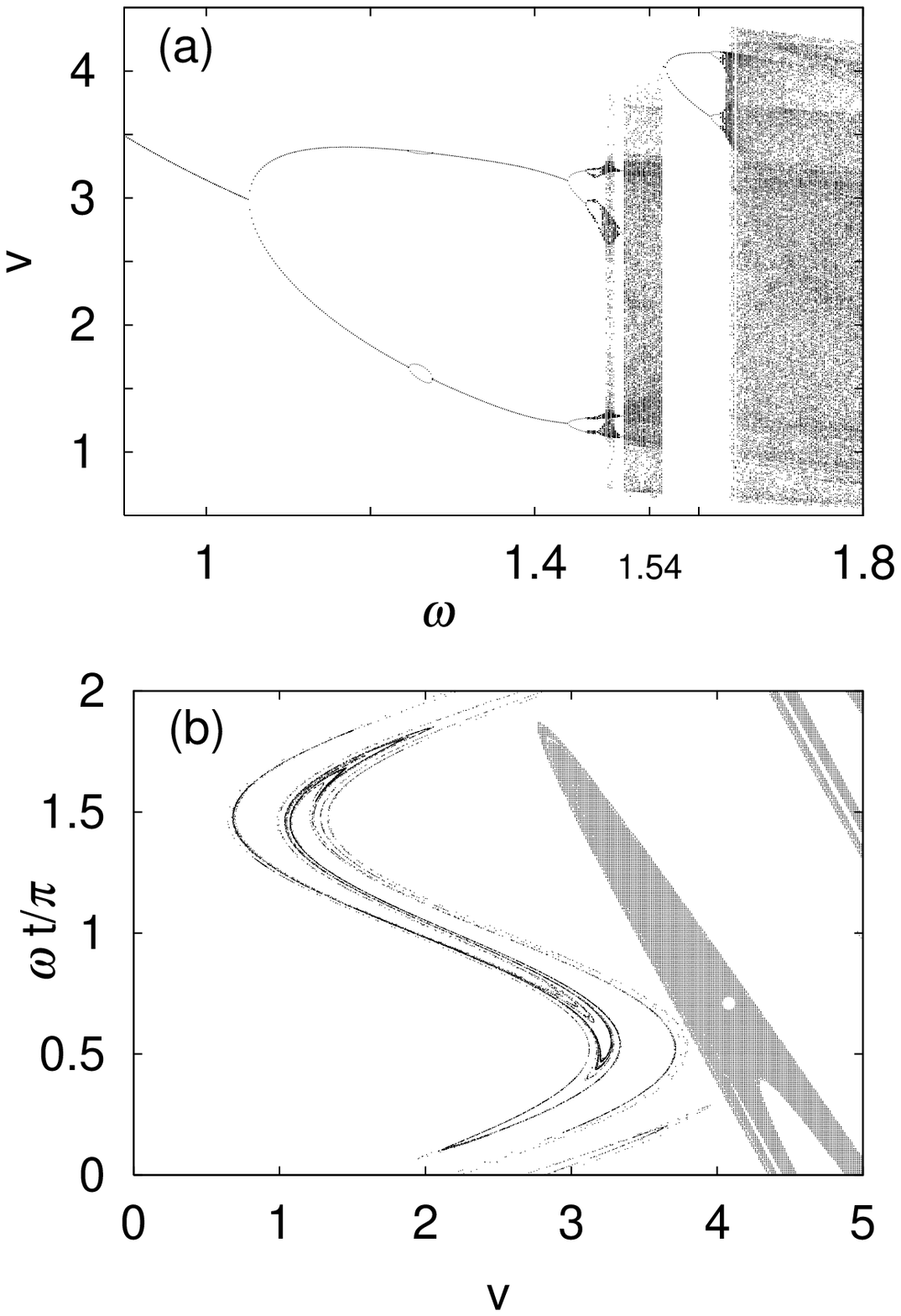}
\end{center}
\vskip 2.5cm
\caption{(a) Bifurcation diagram upon changing $\omega$ with $r=0.39$.
(b) The chaos attractor (dots) and the coexisting period one attractor
 (open circle) with its basin (filled area) for $\omega=1.54$ and $r=0.39$.}
\label{Fig-2}
\end{figure}
\newpage
\centerline{Shohei {\sc Fukano},Yumino {\sc Hayase},and Hiizu {\sc Nakanishi}}
\vskip 3.5cm
\begin{figure}
\begin{center}
\epsfig{width=10cm,file=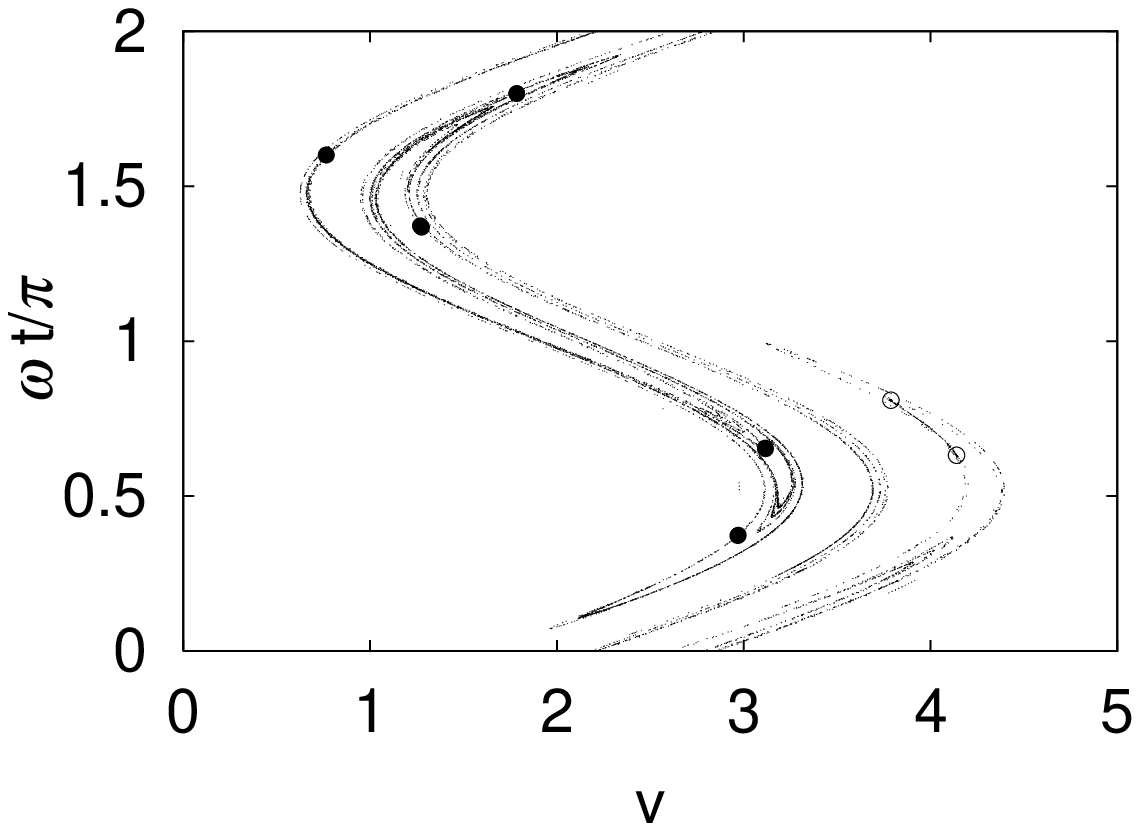}
\end{center}
\vskip 3.5cm \caption{The period ten attractor (solid circles) and the
period one attractor (open circles) with the temporal attractor of the
transient chaos (dots) for $\omega=1.5852$ and $r=0.39$ The orbit of the
period ten attracter consists of five sets of two points, but the two
points within each set cannot be distinguished in the figure because
these two points are very close to each other.  } \label{Fig-3}
\end{figure}
\newpage
\centerline{Shohei {\sc Fukano},Yumino {\sc Hayase},and Hiizu {\sc Nakanishi}}
\vskip 3.5cm
\begin{figure}
\begin{center}
\epsfig{width=10cm,file=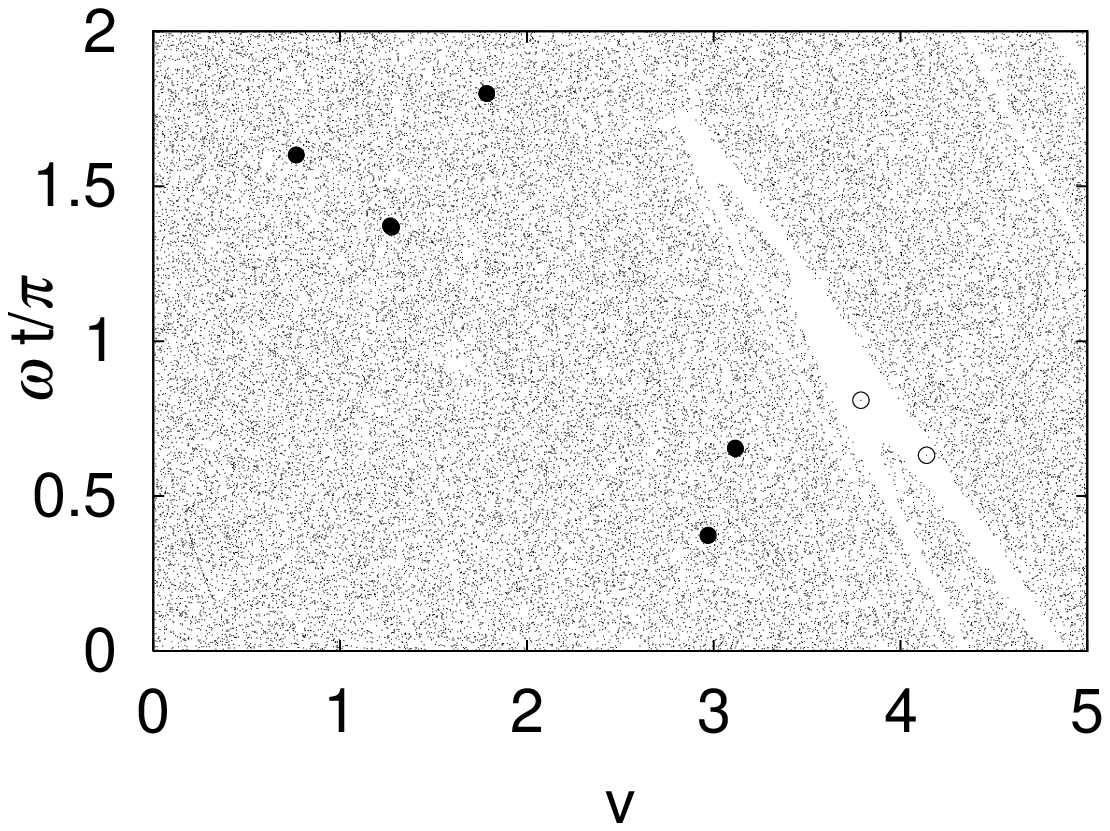}
\end{center}
\vskip 3.5cm
\caption{The basin of the period ten attractor (dots) for 
$\omega = 1.5852$ and $r=0.39$. The attractors of the period ten (solid
 circles) and the period two (open circles) orbits are also indicated.}
\label{Fig-4}
\end{figure}
\newpage
\centerline{Shohei {\sc Fukano},Yumino {\sc Hayase},and Hiizu {\sc Nakanishi}}
\vskip 3.5cm
\begin{figure}
\begin{center}
\epsfig{width=10cm,file=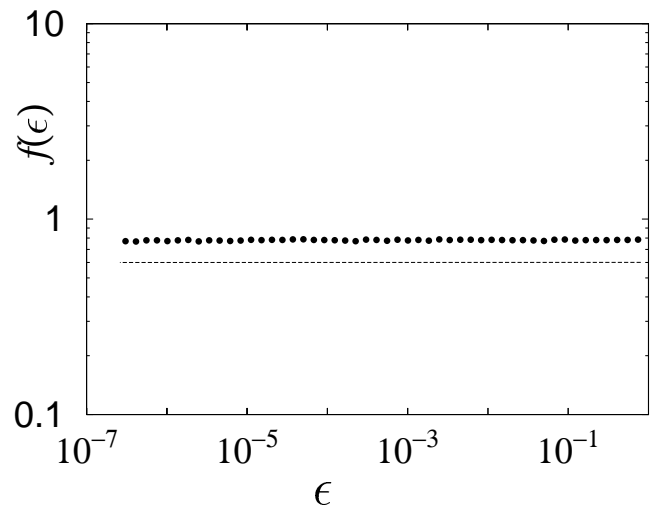}
\end{center}
\vskip 3.5cm \caption{The logarithmic plot of the probability
$f(\epsilon)$ that two points separated by $\epsilon$ belong to
different basins, for $\omega=1.5852$ and $r=0.39$.  The dashed line is
the line with the uncertainty exponent $\alpha=6.2\times 10^{-4}$.} 
\label{Fig-5}
\end{figure}
\newpage

\end{document}